\documentclass[twocolumn, traditabstract]{aa}
\usepackage{color}
\usepackage{soul}
\usepackage{graphicx}
\usepackage{txfonts}
\usepackage{lscape}
\usepackage{longtable}
\sethlcolor{yellow}
\usepackage{natbib}
\def\dg{\mbox{$^\circ\,$}}
\def\arcm{\hbox{$^{\prime}\,\,$}}
\def\arcs{\hbox{$^{\prime\prime}\,$}}
\def\fdg{\mbox{$.\!\!^\circ$}}
\def\farcm{\hbox{$^{\prime}\!\!.\,$}}
\def\farcs{\hbox{$^{\prime\prime}\!\!\!.\,$}}
\def\farcsh{\hbox{$^\mathrm{s}\!\!.\,$}}
\def\hh{$^\mathrm{h}\,$}
\def\mm{$^\mathrm{m}\,$}

\def\mnname{AA/2011/17082}

\begin{document}

\title{VLBI imaging of a flare in the Crab Nebula: More than just a spot.}

\author{A.P.~Lobanov\inst{1} 
	\thanks{Visiting Scientist, University of Hamburg / Deutsches 
Elektronen Synchrotron Forschungszentrum} 
	\and D. Horns\inst{2} 
	\and T.W.B. Muxlow\inst{3} } 

\institute{Max-Planck-Institut f\"ur Radioastronomie, Auf dem H\"ugel
	69, 53121 Bonn, Germany 
	\and Institut f\"ur Experimentalphysik, University of Hamburg, 
	Luruper Chaussee 149, 22761 Hamburg, Germany 
	\and Jodrell Bank Centre for Astrophysics, School of Physics
	and Astronomy, Alan Turing Building, University of Manchester, Manchester, M13 9PL 
}

\abstract{We report on very long baseline interferometry (VLBI)
observations of the radio emission from the inner region of the Crab
Nebula, made at 1.6\,GHz and 5\,GHz after a recent high-energy flare
in this object.  The 5 GHz data have provided only upper limits of
0.4\,milli-Jansky (mJy) on the flux density of the pulsar and
0.4\,mJy/beam on the brightness of the putative flaring region. The
1.6 GHz data have enabled imaging the inner regions of the nebula on
scales of up to $\approx 40\arcs$. The emission from the inner
``wisps'' is detected for the first time with VLBI observations.  A
likely radio counterpart (designated ``C1'') of the putative flaring
region observed with Chandra and HST is detected in the radio image,
with an estimated flux density of $0.5\pm 0.3$\,mJy and a size of
0{\farcs}2--0{\farcs}6. Another compact feature (``C2'') is also
detected in the VLBI image closer to the pulsar, with an estimated
flux density of $0.4\pm0.2$\,mJy and a size smaller than
0{\farcs}2. Combined with the broad-band SED of the flare, the radio
properties of C1 yield a lower limit of $\approx 0.5$\,mG for the
magnetic field and a total minimum energy of $1.2 \times 10^{41}$\,ergs
vested in the flare (corresponding to using about 0.2\% of the pulsar
spin-down power). The 1.6 GHz observations provide upper limits for
the brightness (0.2\,mJy/beam) and total flux density (0.4\,mJy) of
the optical Knot 1 located at 0{\farcs}6 from the pulsar. The absolute
position of the Crab pulsar is determined, and an estimate of the
pulsar proper motion ($\mu_\alpha = -13.0 \pm 0.2$\,mas/yr,
$\mu_\delta = +2.9 \pm 0.1$\,mas/yr) is obtained.  }

\maketitle

\section{Introduction}

The inner region of the Crab Nebula contains a complex and dynamic
structure with a number of elliptical ripples ("wisps") varying in
flux on timescales of days and moving outward with speeds of up to
$0.7\,c$ \citep{tanvir1997,hester2002} in the optical, while radio
measurements with the VLA yield somewhat lower speeds of $\approx
0.3\,c$ \citep{bietenholz2004}. The origin of the wisps is not yet
known, but they are generally thought to be associated with
instabilities downstream of the shock in the wind from the pulsar that
powers the nebula \citep{gallant1994}.

The wind derives its power from the pulsar spin-down injecting an energy
of $\sim 5 \times 10^{38}$\,erg/s into relativistic particles and
magnetic field \citep{rees1974,kennel1984}. The particles are thought
to be mostly electrons and positrons, accelerated to relativistic
energies, with possibly a lesser number of ions \citep{arons1994}.

The exact mechanism by which the spin-down energy is released,
transported, and dissipated remains poorly understood
\citep[cf.,][]{chedia1997,hester1998,begelman1999,lyutikov2003,spitkovsky2004,komissarov2004}
-- with shocks, plasma instabilities, pair acceleration, Poynting
flux, and magnetic field all suggested to affect the evolution of the
wind.

Fast variability possibly related to abrupt energy releases in the
nebula was first reported in the radio on time scales of $\sim 1$ day,
corresponding to an emitting region of $\le 0{\farcs}1$ in size
\citep{matveyenko1975,matveyenko1979}. On 22 September 2010, a
gamma-ray flare was discovered for the first time from the direction
of the Crab Nebula with the AGILE pair-production telescope
\citep{tavani2011}, offering a rare opportunity to follow in detail
the evolution of a flaring region in the nebula. The flare was
confirmed with Fermi/LAT \citep{abdo2011} indicating a variability
time-scale as short as 12~hours \citep{balbo2011}.

The flare was not detected in the keV regime with Swift/XRT
\citep{evangelista2010} and INTEGRAL \citep{ferrigno2010a}. The
Fermi/LAT monitoring had further shown an abrupt decrease of the
gamma-ray flux on 23 September, with the emission returning to its
pre-flare level. Based on the short duration of the high-energy flare,
it has been suggested \citep{komissarov2010} that the flaring
material is located in the optical Knot 1 separated by $\approx 0{\farcs}6$ from the pulsar \citep{hester1995}. However, subsequent
high-resolution imaging with Chandra (on 28 September,
\citeauthor{tennant2010} \citeyear{tennant2010}) revealed three
stationary compact ($\approx 1^{\prime\prime}$ in size) regions
(labelled A,B, and C) of enhanced emission located on the X-ray bright
ring. Similar structures are are also visible in an HST image of the
Crab Nebula taken on 02 October \citep{caraveo2010}.  These knots
present themselves as a viable site for the flaring emission
\citep{tavani2011}, and their compactness may explain naturally the
non-detections in observations with lower spatial resolution where
the extended nebula emission dominates.  The Chandra and HST
observations of the knots indicate a strong, rapid, and efficient
energy release in the flaring material and pose the question of the
relation between the optical, X-ray, and gamma-ray emission.  Given
the position of the knots within the nebula (located not too far off
the jet axis and slightly closer than the innermost wisp), it is not
clear whether the flaring region is associated with the jet or with the
equatorial outflow. Deciding on these issues relies on detailed
imaging of the emitting region, which is best achieved in the radio
regime.

We present here high-resolution wide-field radio images of the flaring
knot and the central region of the nebula, obtained with very long
baseline interferometry (VLBI) observations at a wavelength of
18\,cm. These observations probe a wide range of angular scales, from
$\sim 10$ milliarcseconds to $\sim 40$ arcseconds, detect the flaring
region, and reveal, for the first time, the intricate morphological
structure of the radio emission within about 20 parsecs of the
pulsar. The observation and data reduction are described in
Section~2. The resulting images are presented in Section~3, and discussed
in the context of physics and evolution of the flare in the Crab
Nebula.

\begin{table}
\caption{Antenna parameters} 
\label{tb:antennas}
\begin{center}
\begin{tabular}{l|cccr}\hline\hline
\multicolumn{1}{c|}{Antenna} & \multicolumn{1}{c}{$D$} & \multicolumn{1}{c}{$\lambda_\mathrm{obs}$} & SEFD & \multicolumn{1}{c}{$B_\mathrm{min}$--$B_\mathrm{max}$} \\
  & \multicolumn{1}{c}{[m]} & \multicolumn{1}{c}{[cm]} & \multicolumn{1}{c}{[Jy]} & \multicolumn{1}{c}{[km]} \\ \hline
Effelsberg      & 100     & 6,\,18 &  20,19 &  266--~~853   \\
Westerbork      &  66$^a$ & 6,\,18 &  60,30   &  266--1002  \\
Jodrell Bank    &  25     & 6,\,18 &  320,320 &   17--1388  \\
Medicina        &  32     & 6,\,18 &  170,600 &  757--1429  \\
Onsala          &  25     & 6,\,18 &  600,320 &  601--1429  \\
Torun           &  32     & 6,\,18 &  220,230 &  637--1441  \\
Yebes           &  40     & 6      &  90  & 1313--2152  \\
Cambridge$^b$   &  32     & 6      &  136 &  197--1313  \\
Darnhall$^b$    &  25     & 18     &  356 &   17--1404  \\
Knockin$^b$     &  25     & 18     &  356 &   67--1441  \\
Hartebeesthoeck$^c$ &  26 & 18 &  450 & 7453--8525  \\\hline
\end{tabular}
\end{center}
\small {Notes:}~$a$ -- for Westerbork operating as a phased array; $b$
-- MERLIN antennas; $c$ -- Hartebeesthoeck was added ad hoc in the
observation at 18\,cm, while participating in the e-VLBI session on
November 5. Column designation: $D$ -- antenna diameter; $\lambda_\mathrm{obs}$ -- observing wavelength used; SEFD --
system effective flux density describing the antenna sensitivity at corresponding wavelength;
$B_\mathrm{min}$--$B_\mathrm{max}$ -- range of the telescope
separations from a given antenna (for European telescopes, baselines
to the South African antenna in Hartebeesthoeck are not considered).
\normalsize
\end{table}

\begin{table}
\caption{Target and calibrator sources}
\label{tb:targets}
\begin{center}
\begin{tabular}{clc}\hline\hline
\multicolumn{2}{c}{Object/Coordinates} & $\sigma_\mathrm{pos}$[mas] \\ \hline
\multicolumn{2}{c}{{ Crab-A}$^a$}  &  \\
$\alpha_\mathrm{J2000}$ & ~~05{\hh}34{\mm}32{\farcsh}332 & $\sim$10 \\
$\delta_\mathrm{J2000}$ & +22{\dg}00{\arcm}52{\farcs}45 & $\sim$10 \\
\multicolumn{2}{c}{{J0518+2054}$^b$} & \\
$\alpha_\mathrm{J2000}$ & ~~05{\hh}18{\mm}03{\farcsh}824510 & 0.12\\
$\delta_\mathrm{J2000}$ & +20{\dg}54{\arcm}52{\farcs}49739 & 0.12 \\\hline
\end{tabular}
\end{center}
\small
{Notes:}~$\sigma_\mathrm{pos}$ -- position errors in right ascension and declination. References: $a$ -- \citeauthor{tavani2011} \citeyear{tavani2011} and this work; $b$ -- \citeauthor{petrov2008} \citeyear{petrov2008}.
\normalsize
\end{table}

\section{Observations and data analysis}

The flaring region was observed with VLBI on 5 November 2010 at
1.6\,GHz ($\lambda_\mathrm{obs} = 18$\,cm) and on 23 November 2010 at
4.9\,GHz ($\lambda_\mathrm{obs} = 6$\,cm). The total of eight
EVN\footnote{European VLBI Network, www.evlbi.org} telescopes and
three MERLIN\footnote{Multi Element Radio Linked Interferometer
Network, www.merlin.ac.uk} telescopes participated in the
observations, operating in e-VLBI\footnote{electronic-VLBI, a mode of
VLBI operations, with direct links between participating telescopes
and the correlator maintained via optical fibre connections;
http://services.jive.nl/evlbi/} mode. Basic parameters of the
participating antennas and ranges of the respective baseline lengths,
$B$, are given in Table~\ref{tb:antennas}.

\begin{figure}[t]
\includegraphics[width=0.5\textwidth,clip=true]{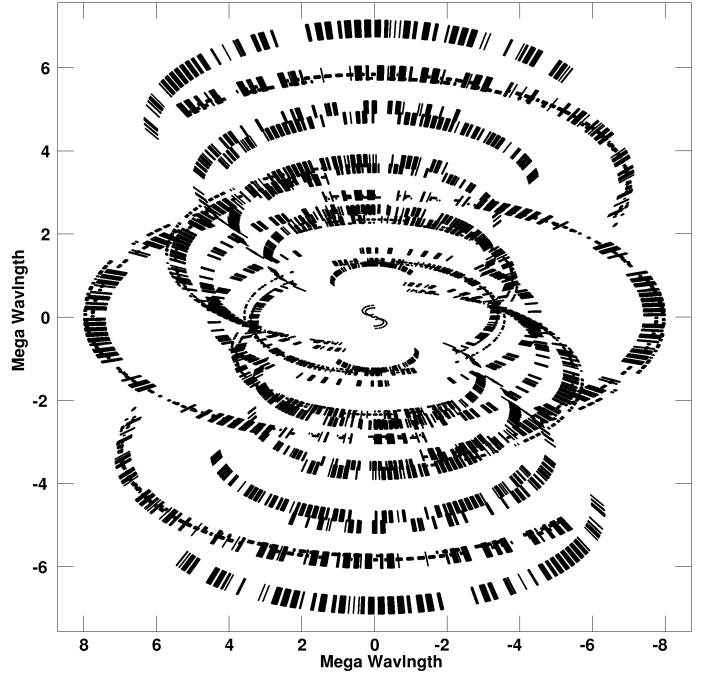}
\caption{Spatial frequency coverage (Fourier domain or {\em
uv}-coverage) of the VLBI visibility data for the target source at
1.6\,GHz. The coverage is plotted in Megawavelength (M$\lambda$)
units, referring to spatial frequencies of
$B_{l,m}/(\lambda_\mathrm{obs}\times 10^6)$, where $B_{l,m}$ is the
projected baseline length for a given pair of antennas
$(l,m)$. Baselines to Hartebeesthoeck, covering a range of
36--48.8\,M$\lambda$, are not shown. The smallest spatial frequency
sampled is $10^{-2}$\,M$\lambda$, corresponding to $\approx$2$\arcs$,
but the central gap in the {\em uv}-coverage results in the structural
sensitivity substantially reduced for scales larger than
$0{\farcs}2$.}
\label{fg:uv}
\end{figure}

The 1.6\,GHz observations were made over 11 hours at a central
reference frequency of $\nu_\mathrm{obs} = 1616.49$\,MHz.  The
4.9\,GHz observations were made over 8.2 hours at a central reference
frequency of $\nu_\mathrm{obs} = 4948.74$\,MHz. The 6\,cm and 18\,cm
observations have sampled ranges of visibility ({\em uv}) spacings,
$B/\lambda$, of 2--$36\times 10^6\,\lambda$
(2--$36\,\mathrm{M}\lambda$) and 0.01--$49\,\mathrm{M}\lambda$,
respectively (note that Hartebeesthoeck took part only in the 18\,cm
observation).

For both observations, the field of view was centered on the position
of the flaring region, Crab-A \citep{tavani2011}, as measured from the
HST data. The nearby bright ($S_\mathrm{1.6GHz} = 0.42$\,Jy) and
compact radio source J0518+2054, separated by 4{\fdg}0 from Crab-A,
was used as a phase-reference calibrator. A calibrator/target cycle of
2/5 minutes was used, with about 30 seconds allocated for slewing. Two
other calibrators, J0530+1331 and 3C\,48, were observed as fringe
finders.  Positions and flux densities of the target and
phase-reference calibrator are given in Table~\ref{tb:targets}.

The EVN data were recorded at a rate of 1024\,Megabit/s (Mbps), in
dual-circular polarisation, which yielded and observing bandwidth of
128\,MHz per polarisation. The total bandwidth was divided into 8
intermediate frequency (IF) bands, each covering 16\,MHz and split in
32 spectral channels of 0.5\,MHz in width. The MERLIN data were
obtained at a 128\,Mbps recoding rate, in dual-circular polarisation,
yielding an observing bandwidth of 16\,MHz per polarisation. The
MERLIN data, corresponding to a single IF band of the EVN data, were
recorded also using 32 spectral channels of 0.5\,MHz in width, for
each polarisation channel. The recording was made in a two-bit
recording mode at all participating antennas.  The resulting coverage
of the Fourier domain ({\em uv}-coverage) provided by the visibility
data on the target source is shown in Fig.~\ref{fg:uv}, for the
observations at 1.6\,GHz.

Correlation of the data was done at the EVN correlator facility of the
Joint Institute for VLBI in Europe (JIVE). The output correlated data
were produced with a 2 seconds averaging time. The combined choice of
the integration time and the frequency channel width ensured that a
field of view as large as 1{\farcm}6 (at 1.6\,GHz) and 0{\farcm}5 (at
5\,GHz) could be imaged using only the visibility data from the
European baselines ($B\le 10\,\mathrm{M}\lambda$) shown in
Fig.~\ref{fg:uv}, with the synthesised beam degraded by less than
10\%. No pulsar gating was applied (this correlation mode is currently
unavailable for e-VLBI observations).

The correlated data were processed using AIPS\footnote{Astronomical
Image Processing Software, National Radio Astronomy Observatory, USA}
software. The data scans were first flagged using automatic flag
tables produced during the correlation. The visibility amplitudes
were calibrated, based on the system temperature measurements and
station gain information. The data for the phase-reference calibrator
were then fringe-fitted, and a bandpass calibration was applied. The
phase-reference calibrator was imaged and found to have only a weak
extended structure, ensuring that it was not contaminating the phase
solutions. The phase solutions were then transferred to the target
scans, and phase-calibrated data for the target source were
imaged.

\subsection{5\,GHz data}

No positive detection of emission can be made from the 5\,GHz data
for the flaring region and for the pulsar itself, even in a
heavily tapered image with a beam (FWHM) of $0{\farcs}12 \times
0{\farcs}03$.  The pulsar emission is reported to have a continuum
flux density of $14.4 \pm 3.2$\,mJy at 1.4\,GHz and a very soft
spectral index $\alpha_\mathrm{psr} = -3.1\pm 0.2$
\citep{lorimer1995}. This corresponds to a flux density of $\approx
0.3$\,mJy at 5\,GHz.  We estimate an r.m.s. noise of
$\sigma_\mathrm{6cm} \approx 0.16$\,mJy/beam, from the inverted 5\,GHz
visibility data. Thus the pulsar flux is expected to reach an $\approx
2\,\sigma$ level in the 5\,GHz image. Based on these considerations,
an upper limit of $\sim 0.4$\,mJy can be provided for the flux density
of the pulsar. The corresponding upper limit on the brightness of the
knot A is then $\sim 0.4$\,mJy/beam, for the above mentioned beam size.

The following discussion is therefore focused primarily on the results
obtained from the VLBI data at 1.6\,GHz.

\subsection{Effects of the pulsar and extended nebula on imaging}

The presence of the pulsar and an extremely large ($420\arcs \times
290\arcs$) and bright nebula results in an increase in the image noise
level in the flaring region.  For the antenna configuration used in
the observation, sensitivity reductions by factors of 3.0 and 3.2 are
expected for the EVN and MERLIN parts of the data, respectively. These
estimates account for the 1.6 GHz flux density ($\sim 900$\,Jy) of the
nebula and the primary beams response of the EVN and MERLIN antennas
to the nebula. The resulting expected r.m.s. image noise in the
combined data is $\approx 60\,\mu$Jy/beam (in a full-resolution image
made from naturally-weighted visibility data).  The contribution of
the pulsar to the image noise is not as significant.  At the observing
frequency of 1.6\,GHz, the expected continuum flux of 8.6\,mJy, and
sidelobes from the pulsar resulting from deconvolution defects are
expected to become relevant at a $\sim$10$\,\mu$Jy/beam brightness
level, which is lower than the estimated thermal noise.

\subsection{Resolution and structural sensitivity}

The visibility data at 18\,cm sample a range of {\em uv}-spacings from
$10^{-2}$\,M$\lambda$ to 48.8\,M$\lambda$, with a large gap between
9\,M$\lambda$ and 36\,M$\lambda$.  The full-resolution, image made
from the uniformly-weighted data, including the baselines to
Hartebeesthoeck, has a synthesised beam with a full width at half
maximum (FWHM) of $19.9 \times 2.2$ mas, while the European baselines
yield a beam with a FWHM of $25.7 \times 22.1$ mas. The
combination of angular scales sampled and flux density sensitivity of
the data on the baselines to Hartebeesthoeck makes them useful only
for imaging and astrometry of the pulsar itself.

The largest formally detectable scale in the data is about 20{\arcs}
($\approx$20\% of the unaberrated field of view). However, this scale
is sampled only by a single baseline, and a realistic limit on
adequate structural sampling is $\approx$$0{\farcs}2$
($\approx$1$\,\mathrm{M}\lambda$), noting the gap in the {\em
uv}-coverage shown in Fig.~\ref{fg:uv}. This gap corresponds to
angular scales of 0{\farcs}2--2\arcs. It should be thus expected that
smooth emission on scales larger than $0{\farcs}2$ (and in the
0{\farcs}2--2\arcs range in particular) could only be poorly recovered
from the data (leading to ``spotty'' structures in images obtained
using deconvolution, or requiring progressively higher signal-to-noise
ratio on these scales in order to be successfully recovered by the
maximum entropy method).  The limited structural sensitivity of the
VLBI data acts effectively as a low frequency {\em uv}-filter,
reducing the contribution from large spatial scales where diffuse
emission from the extended nebula dominates strongly. This leads to a
better detectability of bright and compact regions which otherwise
would have been swamped by the bright diffuse emission. 

\subsection{Imaging}

Taking into account the issues of resolution and structural
sensitivity outlined above, we have made three different images from
the data: {\em 1)} a full-resolution image of a small region around
the pulsar (Fig.~\ref{fg:pulsar}); {\em 2)} a single-scale CLEAN image
aiming at detecting the flaring region (Fig.~\ref{fg:flare}); and {\em
3)}~a multi-scale CLEAN image of extended emission within a region of
about 40$\arcs$ in size (Fig.~\ref{fg:nebula}).  The latter two images
have been made using only the data from European baselines covering
the {\em uv}-spacings of up to 10\,M$\lambda$ (this has been done in
order to improve the sensitivity to extended, low surface brightness
emission). Properties of the images obtained are summarised in
Table~\ref{tb:images}

\begin{table}
\caption{Properties of eVLBI images of the Crab Nebula at 1.6\,GHz}
\label{tb:images}
\begin{center}
\begin{tabular}{lccc}\hline\hline
Image & Beam & $S_{\mathrm peak}$ & $\sigma_\mathrm{rms}$ \\ 
      & [arcsec] & [mJy/beam] & [mJy/beam] \\ \hline
1 (Fig.~2) & $0.020 \times 0.002$ & 8.35 & 0.28    \\
2 (Fig.~3) & $0.150 \times 0.150$ & 8.18 & 0.20   \\
3 (Fig.~4) & $0.500 \times 0.500$ & 8.53 & 0.16    \\ \hline
\end{tabular}
\end{center}
\end{table}

\subsubsection{Pulsar}

\begin{figure}
\includegraphics[width=0.47\textwidth,angle=-90,clip=true]{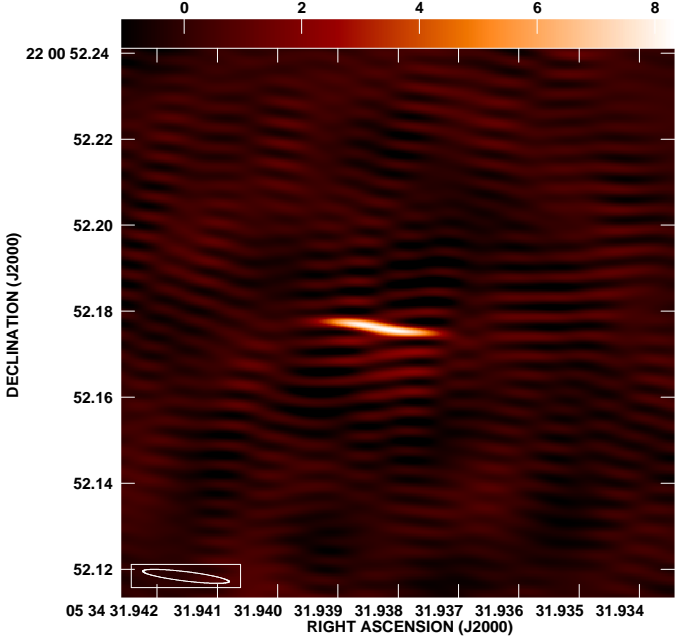}
\caption{Full resolution image of the Crab pulsar obtained from
uniformly weighted data. The image has a peak flux density of
8.3\,mJy/beam and an rms noise of 0.3\,mJy/beam. The synthesised
(restoring) beam is $19.9 \times 2.2$ mas, oriented at a
$\mathrm{P.A.} = 83\dg$. The extreme ellipticity of the restoring beam
results from a very limited East-West {\em uv}-coverage on baselines
$\ge 10\,\mathrm{M}\lambda$.}
\label{fg:pulsar}
\end{figure}

The full-resolution image of the Crab pulsar is made from the entire
data, using the uniform weighting and restricting the field of view to
$0{\farcs}3$. A phase shift of $\Delta\alpha = -5{\farcs}479$ in
right ascension and $\Delta\delta = -0{\farcs}273$ is applied to the
data in order to center the image on the location of the pulsar.  The
resulting high-resolution phase-referenced image of the pulsar is
shown in Fig.~\ref{fg:pulsar}. 

Properties of the pulsar emission are estimated by fitting an
elliptical Gaussian to the image brightness distribution. This yields
an integrated flux density of $8.4 \pm 0.5$ mJy, which is
close to the value estimated from the flux density measurements of
\cite{lorimer1995}. Deconvolution of the fitted extent of the Gaussian
indicates that the pulsar emission is not resolved, with the limits on
its size obtained from the uncertainties of the fit
$\sigma_\mathrm{maj} = 0.40$\,mas (at a position angle of $81\dg$) and
$\sigma_\mathrm{min} = 0.09$\,mas for the major and minor axis of the
Gaussian, respectively. A similar estimate ($\sigma_\mathrm{maj} =
0.40$\,mas, $\sigma_\mathrm{min} = 0.04$\,mas) is obtained using the
SNR-based resolution criterion \citep{lobanov2005}.

\subsubsection{Image of the flaring region}

As described above, extended emission of the Crab nebula has
been imaged with the data from the European baselines ($B\le
10\,\mathrm{M}\lambda$), which has been done in order to improve the
sensitivity to emission on scales $\gtrsim 20$\,mas resolved out on
the baselines to Hartebeesthoeck.

The visibility data are phase-shifted in right ascension by
$\Delta_\alpha = -6{\farcs}5$, in order to fit better the field of
interest within a rectangular grid. The grid extends by $20{\farcs}9$
from the phase-center, thus covering almost the entire unaberrated
field of view (allowing for about 5\% reduction of the peak response
due to bandwidth and time average smearing). As the flaring region is
likely to be extended, the visibility data are weighted using the
natural weighting and tapered using a Gaussian taper with
$\sigma_\mathrm{taper} = 2.2\,\mathrm{M}\lambda$. This yields a
synthesised beam of $54~\mathrm{mas} \times 49~\mathrm{mas}$, which is
about 2.5 times larger than the beam obtained without the taper.

\begin{figure}
\includegraphics[width=0.47\textwidth,angle=-90,clip=true]{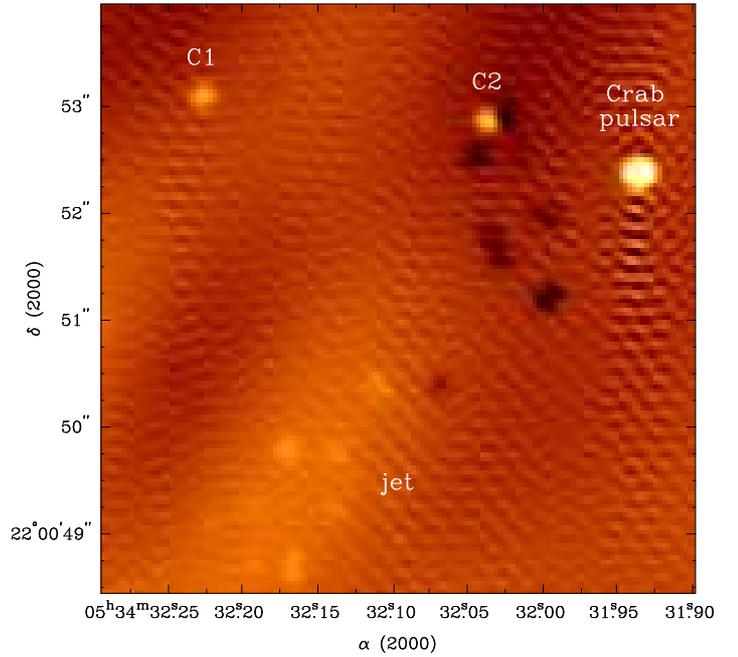}
\caption{Image of the pulsar and the flaring region. The image is
obtained by applying a single-scale CLEAN deconvolution on the {\em
uv}-tapered visibility data on the European baselines. The parameters
of the image are given in Table~\ref{tb:images}. For the presentation
purpose the image is convolved with a $0\farcs 15$ beam (three times
the natural restoring beam).}
\label{fg:flare}
\end{figure}

The image of the flaring region shown in Fig.~\ref{fg:flare} is
obtained by using single-scale CLEAN deconvolution and convolving the
results with a larger beam ($0\farcs 15$) in order to enhance the
contrast for the regions of weak, extended emission. The likely
radio counterpart of the flaring component (denoted as C1) is clearly
detected in this image, albeit at a somewhat smaller separation from
the pulsar than the optical/X-ray emitting region Crab-A. The weak
radio jet (located in the direction of the ``sprite'' identified in
the HST images; {\em cf.}, \citeauthor{hester2002}
\citeyear{hester2002}; \citeauthor{melatos2005}
\citeyear{melatos2005}) and another, weaker but more compact
component (C2) are also visible in the image. 
The possible
nature of the two compact components and their relation to the flare
will be discussed below.

The bright optical knot (Knot 1) seen in the HST images at a
0{\farcs}6 distance from the pulsar \citep[{\em cf.},][]{hester1995} is not
detected in the 1.6\,GHz image. An upper limit on its brightness is
provided by the rms (0.2\,mJy/beam) measured in the region where it is
located. The non-detection of Knot 1 may imply both its weakness in
the radio and large extension resulting in it being resolved out in
the VLBI data. For the Knot 1 dimensions ($0{\farcs}50 \times
0{\farcs}16$, \citeauthor{hester1995} \citeyear{hester1995}), the
upper limit on its total flux density at 1.6\,GHz is $\approx
0.4$\,mJy.

\begin{figure*}[t!]
\includegraphics[width=0.9\textwidth,angle=-90,clip=true]{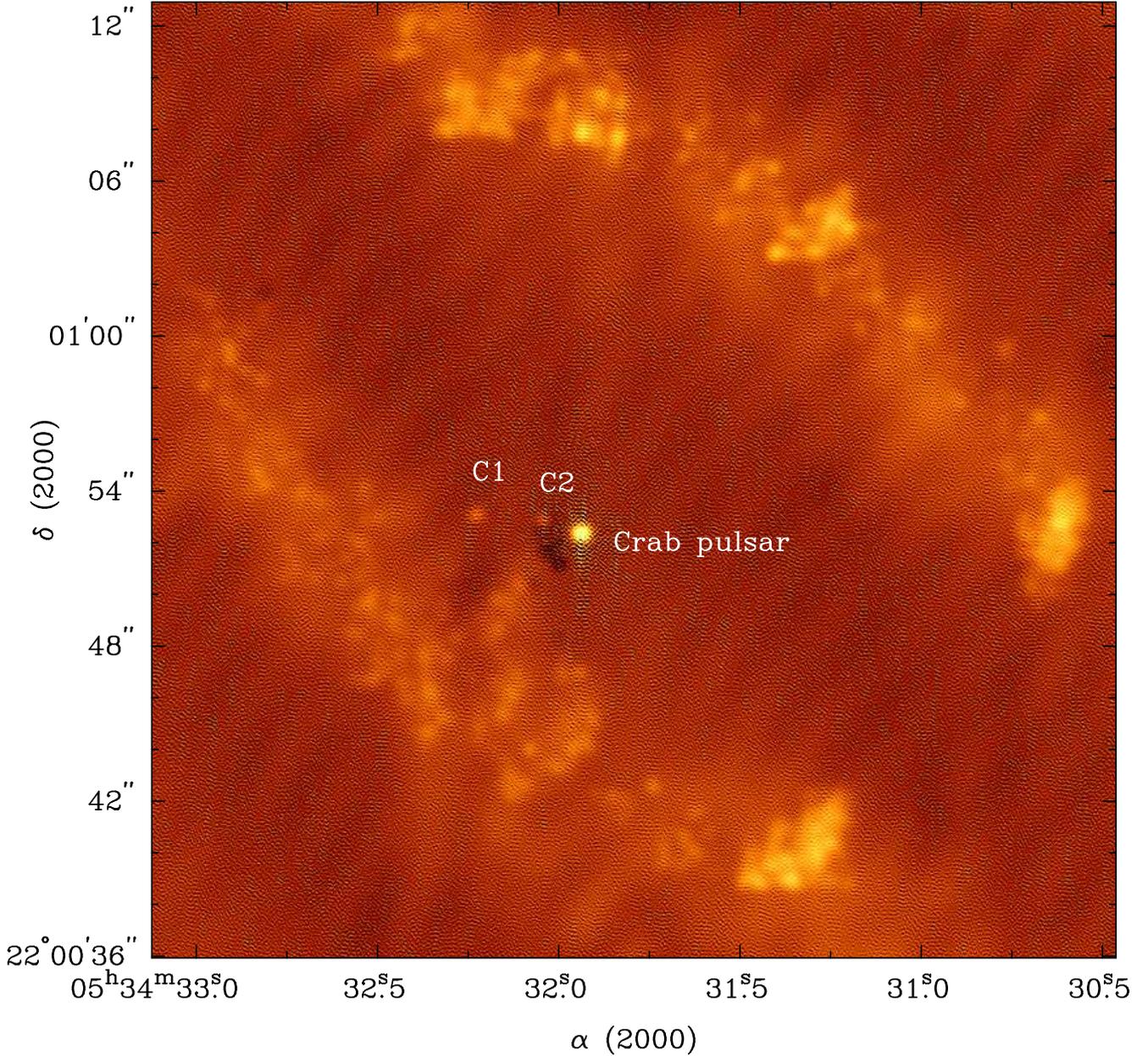}
\caption{Image of the central region of the Crab nebula obtained from
naturally weighted data, {\em uv}-tapered data on baselines up to
10\,M$\lambda$. The image is obtained using a multi-scale CLEAN
deconvolution and restored with a circular beam of $0{\farcs}5$ in
size. The image has a peak flux density of 8.5\,mJy/beam and an rms
noise of 0.16\,mJy/beam. The total flux density recovered in the image
is 148\,mJy.}
\label{fg:nebula}
\end{figure*}

\subsection{Fidelity of detection of the compact features}

The compact components C1 and C2 are strongly resolved, and they
cannot be detected without tapering of the visibility data. This
implies that their sizes should be {\em a priori} larger than $\sim
40$\,mas and that both emitting regions most likely have a smooth
extended structure that can be adequately imaged only with a dense
sampling on short baselines ($B\le 250$\,km). Because of the lack of
such baselines in our data, the appearance and relative brightness of
C1 and C2 depend on the choice of deconvolution and image restoration
procedures. In the single-scale CLEAN image in Fig.~\ref{fg:flare},
which is biased to higher spatial frequencies (hence smaller angular
scales), the peak brightness is higher in the more compact feature
C2. The multi-scale CLEAN image in Fig.~\ref{fg:nebula} accounts
better for lower spatial frequencies (larger angular scales), yielding
a larger flux density for the more extended feature C1, with some of
this emission possibly coming from the extended emission of the nearby
``wisp''. Considering all these factors, and taking into account the
likely angular extent of both components, we estimate that the
application of single-scale CLEAN deconvolution yields detections at
an SNR of 4.2 and 3.2, for C1 and C2 respectively. The respective SNR
values are reduced to 3.8 and 2.5 in the multi-scale CLEAN image in which
the cumulative spatial sensitivity is shifted to scales that are much larger
than the angular extent of C1 and C2.

To test the fidelity of detection of such features in the EVN data, we
amend the visibility data with two sets of simulated Gaussian
components representing C1 and C2 at different position angles and
distances from the pulsar. The representations of C1 have a flux
density of 0.5\,mJy and a size of 0{\farcs}2; the respective values
for C2 are 0.4\,mJy and 0{\farcs}15. Random noise is added to
the visibility data, to accommodate for the total signal increase due
to the simulated features. As the sum flux density of the simulated
components (1.8\,mJy) represents only a small fraction of the total
flux density in the image (140\,mJy), the resulting noise factor is
also relatively small and leads to an increase of the average
visibility noise from 73.2\,mJy to 74.1\,mJy.

Locations of the first set (C1a, C2a) of the added components are
obtained by rotation of the positions of C1 and C2 by $-122\dg$. For
the second set (C1b, C2b), a rotation by $-47\dg$ is combined with a
factor of 4 stretch in distance from the pulsar. This ensures that the
resulting locations of the simulated components span a range of pulsar
separations, $\Delta r$, ranging from 1{\farcs}6 to 16{\farcs}4 and
fall outside of the areas with detected emission. The modified data
are then imaged using the same procedure as applied for obtaining
the single-scale CLEAN image. All four simulated features are detected
in the resulting image and the ratios, $\zeta_\mathrm{S}$, of their
fluxes to the fluxes of the original components are close to unity
(see Table~\ref{tb:ratios}), and only the most distant simulated feature
shows a perceivable decrease in flux density, owing most likely to the
smearing effects. These results support further the reliability of the
detection of the compact features C1 and C2.

\begin{table}[ht]
\caption{Flux density ratios for simulated compact features}
\label{tb:ratios}
\begin{center}
\begin{tabular}{c|rrc}\hline\hline
Name &  \multicolumn{1}{c}{$\Delta r$} & \multicolumn{1}{c}{P.A.} & $\zeta_\mathrm{S}$ \\ \hline
C2a  &  1{\farcs}6 & -49{\fdg}0 & $1.11 \pm 0.51$      \\
C1a  &  4{\farcs}1 & -41{\fdg}0 & $0.91 \pm 0.32$      \\
C2b  &  6{\farcs}6 &  26{\fdg}0 & $0.96 \pm 0.47$      \\
C1a  & 16{\farcs}4 &  34{\fdg}0 & $0.79 \pm 0.31$      \\ \hline
\end{tabular}
\end{center}
Notes: $\Delta r$, P.A. -- component distance and position angle with
respect to the pulsar; $\zeta_\mathrm{S}$ -- measured ratio between
the flux density obtained in the test image for the simulated and the
original feature.
\end{table}

\subsubsection{Image of the central region of the nebula}

Beyond the extent of the flaring region, the CLEAN algorithm finds
positive flux as well, indicating likely detection of the two inner
``wisps'' visible in the optical and radio images obtained with HST
and VLA \citep{bietenholz2004}. As expected, this emission is not well
recovered by CLEAN, resulting in a ``spotty'' appearance of the
elliptically-shaped wisps.  In order to improve the representation of
large-scale structures in the image, we have applied 
multi-scale CLEAN (MS-CLEAN) algorithm.

\begin{figure}[t]
\includegraphics[width=0.47\textwidth,angle=-0]{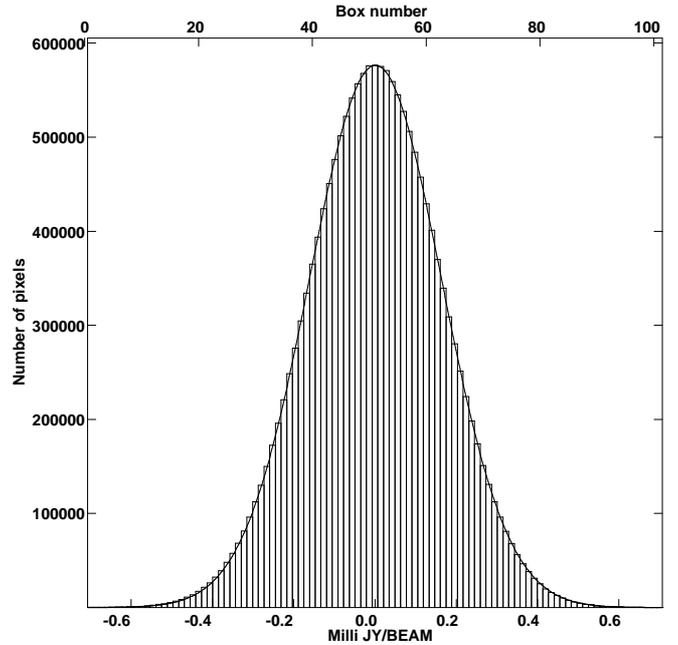}
\caption{Histogram of the residual flux distribution in the VLBI
image, after applying the multi-scale CLEAN. The histogram is well
approximated by the Gaussian noise (solid line) with an r.m.s. of
$0.16$\,mJy/beam.}
\label{fg:noise}
\end{figure}

Application of MS-CLEAN leads to moderate improvement of
the image at large scales (Fig.~\ref{fg:nebula}), and it has also
resulted in a moderate reduction of the r.m.s. noise in the image.
The distribution of the residual flux in the image is shown in
Fig.~\ref{fg:noise}. The distribution is essentially Gaussian,
implying an r.m.s. of $0.16$\,mJy/beam.
At the same time, the application of multi-scale CLEAN has introduced
a substantial power on scales $\ge 0\farcs 5$ and thus reducing the
contrast on smaller scales (which is visible in particularly in
Fig.~\ref{fg:nebula} at the location of the weaker feature C2).

The total CLEAN flux density in the image is 148\,mJy, which reflects
only a fraction of the total flux density in this area (as the VLBI
data do not sample spatial scales larger than 20\arcs and indeed have
problems with recovering flux on spatial scales larger than about
0{\farcs}2.

\subsubsection{Fidelity of the large-scale structure}

\begin{figure}[t]
\includegraphics[width=0.47\textwidth,angle=-90,clip=true]{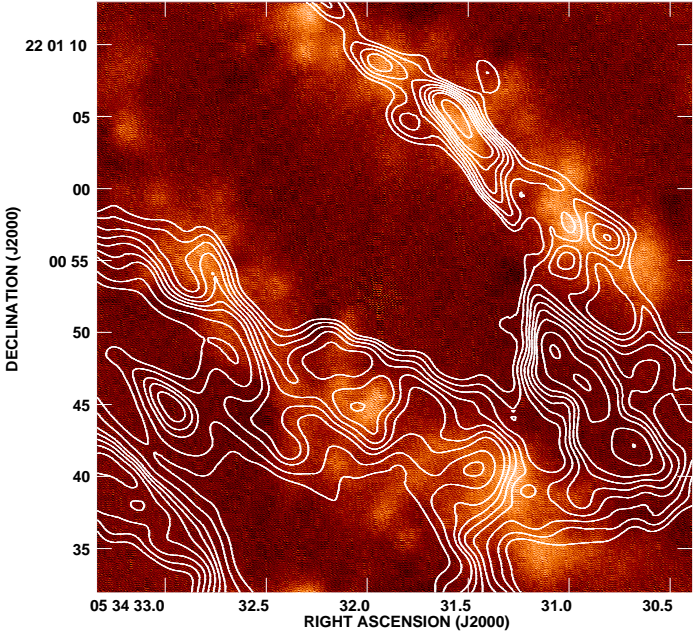}
\caption{The VLBI image of the Crab nebula (grey scale) overlaid with
a VLA image (contours) of the same region obtained by
\cite{bietenholz2004}. The VLBI image is obtained from the
multi-scale CLEAN image by removing the contribution from the pulsar
and restoring the resulting image with a 1{\farcs}4 beam in order to
match the resolution of the VLA image.  Locations of the inner
``wisps'' agree well in the two images, despite potential
complications due to the non-contemporary observing epochs and
possible image artifacts in the VLBI image. See further discussion in
the text.}
\label{fg:evn-vla}
\end{figure}

The large-scale structure detected in the image shown in
Fig.~\ref{fg:nebula} can be compared to the structure imaged with the
VLA \citep{bietenholz2004}. This comparison is presented in
Fig~\ref{fg:evn-vla}, which relates the VLA image to the
multi-scale CLEAN image restored with a 1{\farcs}4 beam (to match the
VLA image resolution). The location and shape of the inner ``wisps''
agree well in the VLBI and VLA images. It should be noted of course
that the two images are not contemporary, and thus the apparent
agreement may be simply fortuitous. Nevertheless, this argues for the
overall plausibility of the larger scale structure in the VLBI image.

One particularly remarkable difference between the structure seen in
the VLA and VLBI images is observed in the south-western direction,
where the VLBI image does not show any enhancement in flux, while a
strong and extended emission region is visible in the VLA image. It is
presently impossible to judge whether this disagreement is caused by
the evolution of the radio emission or by deficiencies in the VLBI
data and imaging. It should be noted however that this region is
situated at more than 30\arcs ($\approx$600 beamwidths) away from the
phase-center of the VLBI data, and the resulting reduction in peak
response and beam deterioration may reduce detectability of weak and
extended emission.

The imaging results summarised above indicate that the
VLBI observations have not only been sensitive enough
to detect the presence of a compact flaring emission in the Crab
Nebula, but they also enable obtaining a detectable response from the
inner wisps of the nebula.  However, the fidelity of localisation of
this emission remains marginal, given the limited {\em uv}-coverage of
the VLBI data.  It is obvious that improved {\em uv}-coverage on
shorter baselines is required in order to be able to image these
structures, and a full-fledged combination of eMERLIN and EVN, both
operating at 1 Gbit/s recording rates, would provide substantially
better images of the central region in the Crab nebula.

The measured brightness of the wisps at
1.6\,GHz and the r.m.s. estimate at 5\,GHz can be reconciled with the
``canonical'' spectral index of $\alpha =- 0.3$ obtained in the radio
regime for the total flux density of the nebula
\citep{baars1977,kovalenko1994} as well as for the resolved structures
including the inner wisps \citep{bietenholz1997}. The inner wisps
detected in the VLBI image at 1.6\,GHz have an average brightness of
0.3\,mJy/beam. Adopting $\alpha = -0.3$ yields an expected 5\,GHz
brightness of 0.2\,mJy/beam, which compares well with the r.m.s. of
0.16\,mJy/beam estimated from the 5\,GHz data.


\section{Results and Discussion}

The VLBI observations of the Crab nebula have yielded accurate
positional and photometric information about the pulsar itself and the
flaring region, as well as marginal morphological and photometric
information about the inner wisps of the nebula.

\subsection{Astrometry and proper motion of the pulsar}

The position of the pulsar is measured relative to the position of the
phase-reference calibrator, J0518+2054. The position of J0518+2054 has
been determined with an accuracy of 0.12 mas, both in right ascension
and declination \citep{petrov2008}. The calibrator image obtained from
the VLBI data shows no significant structure that could affect the
calibrator phases and introduce additional positional errors. The
total flux density recovered in the calibrator image is $290.1 \pm
1.8$\,mJy, and the peak flux density is $271.0 \pm 1.0$\,mJy/beam,
further underlying the compactness of the structure. The calibrator
image has a convolving beam of $18 \times 12$\,mas and an r.m.s. noise
of 0.8\,mJy/beam (and the respective peak-to-noise SNR of 340), thus
limiting the angular size of the calibrator to $\le 1.1$\,mas and
introducing an error of $0.06$\,mas to the positional measurements
\citep{lobanov2005}.  For the calibrator, we obtain the following
J2000 position (the errors listed are formal errors of the fit):
\[
\alpha_\mathrm{J2000} = \textrm{05:18:03.8245014} \pm 0.0000014\,,
\]
\[
\delta_\mathrm{J2000} = \textrm{20:54:52.497662} \pm 0.000028\,.
\]
Analysis of the high-resolution
image of the pulsar yields an apparent J2000 position
(the errors are formal errors of the fit):
\[
\alpha_\mathrm{J2000} = \textrm{05:34:31.9383014} \pm 0.0000081\,,
\]
\[
\delta_\mathrm{J2000} = \textrm{22:00:52.17577} \pm 0.00018\,.
\]
Taking into account the positional errors of the calibrator listed in
Table~\ref{tb:targets}, more conservative estimates of the errors
of the pulsar position are given by $\Delta_\alpha = 0.18$\,mas
(0{\farcsh}000012) and $\Delta_\delta = 0.22$\,mas. These estimates do
not include errors due tropospheric and ionospheric delays and errors
from the calibrator-target phase transfer normally considered in
measurements made with dedicated VLBI astrometry observations
\citep[cf.,][]{guirado1995,martividal2008}. Based on the analysis of
these factors \citep{guirado1995}, we can expect that their
contribution to the positional errors should be within $\approx 0.05\,
\mathrm{FWHM\,[mas]}\,\Delta\theta\mathrm{[deg]}$ milliarcseconds,
with $\Delta\theta$ giving the calibrator-target distance. For the FWHM of
the full resolution image, one can therefore expect that the positional errors
can reach $\Delta\alpha = 3.4$\,mas and $\Delta\delta = 0.45$\,mas (reflecting a much better north-south resolution of the image).

\begin{figure*}[t!]
\includegraphics[height=1.0\textwidth,angle=-90,clip=true]{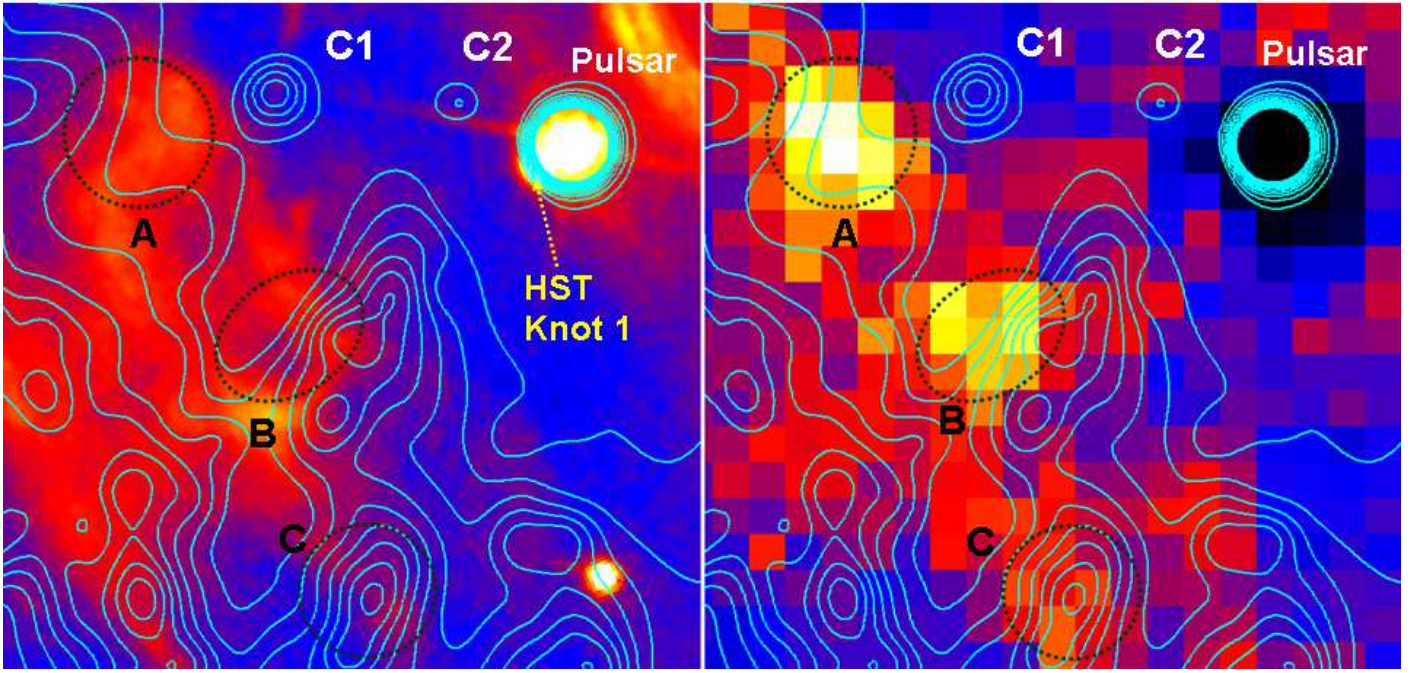}
\caption{Image of the radio emission (green contours; multi-scale
CLEAN image, convolved with a 0{\farcs}7 beam; contours start at
$\mathrm{SNR} = 2$ and increase with an interval of 0.5) overlaid
with images of the optical (left panel, 02 October 2010;
\citeauthor{caraveo2010} \citeyear{caraveo2010}) and X-ray (right
panel, 28 October 2010; \citeauthor{ferrigno2010b}
\citeyear{ferrigno2010b}) emission in the flaring region. The radio
emission from the flaring region C1 is offset from the knot A
\citep{tavani2011} identified in the optical and X-ray emission. The
overall appearance of the optical and radio emission is similar in the
flaring, with a clear ``knee-like'' structure seen in the optical and
trailed by a similar structure in the radio. It should also be
noted that the apparent discrepancy in appearance of C1 and C2 in this
figure and in the image in Fig.~3 (convolved with a $0{\farcs}15$
beam) results from the fact that the component C2 is compact ($\le
0{\farcs}2$ in size) while the size of C1 can be as large as
$0{\farcs}6$ (see discussion in Sect.~\ref{sc:flreg})}
\label{fg:evn-hst}
\end{figure*}

The position measurement made from the VLBI data can be combined with
the Crab pulsar positions obtained from pulsar timing
\citep{taylor1993,han1999}, yielding the following estimate of the
pulsar proper motion:

\[
\mu_\alpha = -13.0 \pm 0.2\quad \textrm{mas/yr}\,,
\]
\[
\mu_\delta = +2.9 \pm 0.1\quad \textrm{mas/yr}\,.
\]
This estimate agrees well with recent estimates made from a long range
of HST observations \citep{ng2006,kaplan2008}.

\subsection{Flaring region}
\label{sc:flreg}

The flaring components C1 and C2 are both very weak, detected at
$\textrm{SNR}\le 4$, which makes it difficult to provide robust
estimates of their properties. Table~\ref{tb:flare} lists our best
estimates of the positions, flux densities, and sizes of the two
components, based on Gaussian fits to the single-scale and multi-scale
image (with the latter providing a somewhat better account of the
extent of the emitting regions). We use these positions to estimate
brightness temperature of the emission in respective regions. The
resulting values indicate brightness temperature in excess of $\approx
4000$\,K (it should be noted here that estimating such low values of
brightness temperature is enabled by the presence of short baselines
provided by the MERLIN antennas). The upper limit on the size of C1
corresponds to the maximum size of a 0.5\,mJy feature that could be
detected in the data.

\begin{table}[ht]
\caption{Properties of compact flaring components}
\label{tb:flare}
\begin{center}
\begin{tabular}{l|c|c}\hline\hline
 & C1 & C2 \\\hline
R.A. [s]      & $32.2271 \pm 0.0011$ & $32.03851\pm 0.00052$ \\
Dec. [\arcs]  & $~53.096  \pm 0.014$ & $~52.8600 \pm 0.0070$ \\
$S_\mathrm{1.6\,GHz}$ [mJy] & $0.5 \pm 0.3$ & $0.4 \pm 0.2$ \\
$\theta$ [\arcs] & 0{\farcs}2 -- 0{\farcs}6 & $\le 0{\farcs}2$ \\
$T_\mathrm{b}$ [K] & 1000 -- 5000 & $\ge 4000$  \\ \hline
\end{tabular}
\end{center}
Notes: $S_\mathrm{1.GHz}$ -- component flux density; $\theta$ --
estimated size (C2 is unresolved, and a limit on the size is estimated
from the SNR following \cite{lobanov2005}; for C1, the upper limit on
the size is given by the detection limit in the data); $T_\mathrm{b}$
-- brightness temperature calculated for the respective size and flux
estimates (note that short baselines of MERLIN
($10^{-2}$--$10^{-1}\mathrm{M}\lambda$) provide sensitivity to
emission with brightness temperature greater than $\approx 1000$\,K).
\end{table} 

An overlay of the radio image of the flaring region with an HST image
from 2 October \citep{caraveo2010} and a Chandra image from 28 October
\citep{ferrigno2010b} is shown in Fig.~\ref{fg:evn-hst}. The radio
image is convolved with a 0{\farcs}7 beam to further emphasise weak
extended emission. This overlay shows that the radio emission
traces well the inner wisp and ``sprite'' (region B), commonly
identified in the optical/X-ray images. The radio component C1 is
located close to a ``knee-like'' region (region A) detected in the
optical and X-ray images and believed to be associated with the flare
\citep{tavani2011}. {This morphology may also be reflected by the
radio emission, but establishing this relation clearly requires better
quality of radio imaging on arcsecond scales.}

\begin{figure*}[t!]
 \includegraphics[height=0.95\textwidth,angle=-90]{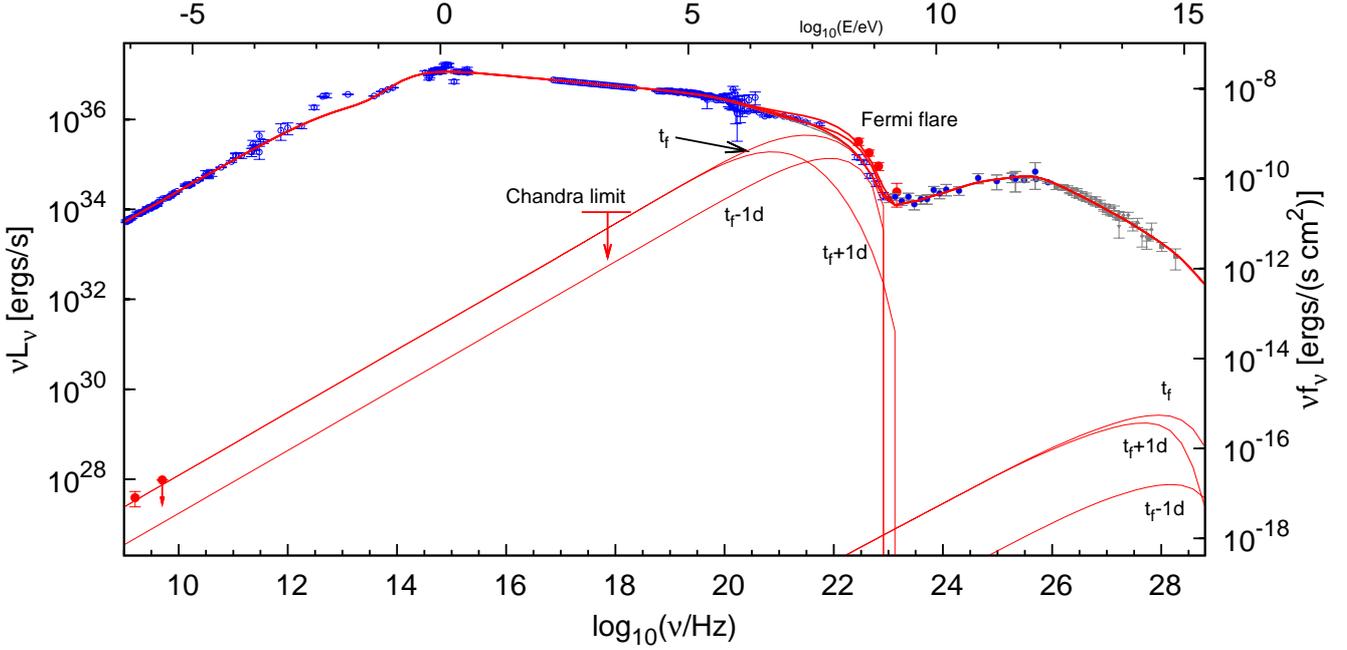}
\caption{\label{fig:SED} The spectral energy distribution of the
nebula and the flaring component C1.  The data on the nebula have been
compiled in \cite{meyer2010}. Details on the modelling of the
synchrotron nebula are provided in that reference as well. In addition
to the nebula, the SED of the feature C1 is shown together with a
time-dependent model of the flare (solid red line), with spectra
calculated at the epoch of the flare maximum ($t_\mathrm{f}$) and one
day before ($t_\mathrm{f-1d}$) and after ($t_\mathrm{f+1d}$) the
maximum }
\end{figure*}

\subsection{Origin of the radio knots}

A significant offset of both C1 and C2 from the jet axis is
evident. If these two features are associated with the jet (and are
not part of the wisps produced in the equatorial outflow), it may be
speculated that they are located on a ``wall'' formed by interaction
of the rotating outflow with the ambient material in the nebula.

The offsets $\Delta\,r$ of C1 and C2 with respect to the jet axis,
$z$, directed at a position angle $\psi_\mathrm{jet} \approx 130\dg$
(estimated from the image in Fig.~\ref{fg:flare}) can be represented
by a power-law expansion $\Delta\,r \propto z^m$, which yields $m
\approx 0.7$. A jet viewing angle of $60\dg$
\citep{cheng2000,komissarov2010} is assumed in these calculations. The
expansion derived is consistent with the one expected for the jet
boundary in a magnetically confined flow \citep{fendt1997,fendt2001}
with a weakly evolving poloidal magnetic field component. One can
assume that the jet boundary is defined by the outermost flux
surface \citep[cf.,][]{fendt2001} and that the collimation of
the flow begins near the light cylinder, $R_\mathrm{lc}$, with an
initial radius of the flow of $\approx 10\,R_\mathrm{lc}$. It would
then require an electric current $I_\mathrm{p} \approx 4 \times
10^{14}\,\mathrm{A}$, a magnetic field $B \approx 4\times 10^{12}$\,G,
and a magnetic flux $\Psi_\mathrm{p} \approx 6 \times
10^{24}\,\mathrm{G}\,\mathrm{cm}^2$ (all agreeing well with earlier
estimates made for the Crab pulsar; cf., \citeauthor{hankins2007}
\citeyear{hankins2007}; \citeauthor{hirotani2006}
\citeyear{hirotani2006}) to explain the locations of the knots C1 and
C2, assuming that these knots are located near the jet
boundary. We note that the value of $I_\mathrm{p}$ is larger than the
Goldreich-Julian current, which can be explained by pair cascades
yielding particle densities well in excess of the Goldreich-Julian
density \citep{arendt2002}. With the jet expansion derived, the jet
should have local half-opening angles of $\phi_\mathrm{C1} = 27\dg$
and $\phi_\mathrm{C2} = 37\dg$ at the locations of the radio
knots. For the HST Knot 1, the respective half-opening angle is
$\phi_\mathrm{HST\,1} = 56\dg$, which is close to the value of $60\dg$
expected for this region \citep{komissarov2010}. These arguments lend
further support to the suggestion that the radio knots are connected
to the outflow.  However, providing a firm conclusion on this matter
would certainly require a closer followup of another prominent flare
in the Crab nebula.

\subsection{Connection to the high-energy flare}

Based on the observational results summarised above, the faint and
compact radio features C1 and C2 could in principle be related to the
gamma-ray flare, even if they are not co-spatial with the active
regions observed in the optical and X-ray bands. In the following, we
consider a scenario in which the brighter feature C1 is related to the
gamma-ray flare and discuss the consequences.

The spectral energy distribution (SED) of C1 is shown in
Fig.~\ref{fig:SED} together with the nebula emission \citep[see][for
details of the data]{meyer2010}. This SED combines the flux
density measured at 1.6\,GHz, the upper limit obtained at 5\,GHz, and
the X-ray flux of the knot A treated here as a (soft) upper limit on
the X-ray emission from the immediate vicinity of C1. Assuming that
the underlying particle distribution present in the region C1 has the
same spectral slope ($\mathrm{d} n/\mathrm{d}\gamma = n_0
\gamma^{-1.60}$) as the electrons responsible for radio emission in
the nebula, the resulting spectrum matches quite well both the radio
as well as the gamma-ray emission across 14 orders of magnitude in
energy.

The observed variability time-scale of about 24~hrs \citep{balbo2011}
constrains the size of the emitting region to be smaller than $r <
t_\mathrm{var} c= 3\times 10^{15}\,\mathrm{cm}\,
(t_\mathrm{var}/10^5\,\mathrm{sec})=
0{\farcs}1\,(t_\mathrm{var}/10^5\,\mathrm{sec})$. This is consistent
with the observed extension of C1.  Assuming that the gyro-radius
$r_\mathrm{g} < ct_\mathrm{var}$ and that gamma-rays are produced via
synchrotron emission, a lower limit on the B-field is set by $B\ge
0.5\,\mathrm{mG}\,(E/100\,\mathrm{MeV})^{1/3}(t_\mathrm{var}/100\,\mathrm{ksec})^{-2/3}$.
A conservative upper limit on the magnetic field of $\approx 100$\,mG
follows from the argument that the total energy in the magnetic field
should not exceed the spin down power of the pulsar integrated over
the duration of the flare.  The equipartition magnetic field (assuming
minimum total energy) is
$B_\mathrm{eq}=3\,\mathrm{mG}\,(t_\mathrm{var}/100\,\mathrm{ksec})^{-6/7}
(E/100\,\mathrm{MeV})^{1/7}$. The corresponding total minimum energy
($w_\mathrm{tot} = w_\mathrm{e} + w_\mathrm{B} = 1.2\times
10^{41}$\,ergs).  The injection power required to generate the burst
is $\dot w\approx w_\mathrm{tot}/t_\mathrm{var}=1.2\times
10^{36}\,\mathrm{ergs/s} = 0.2$\,\% of the available spin-down power.
Even though this is a comfortable margin, deviations from the minimum
energy requirement are limited to roughly one order of magnitude.

Given the limited knowledge about the dynamics of the feature C1, it
is difficult to estimate directly any effects related to relativistic
beaming. A kinematic estimate of the Doppler factor
$\mathcal{D}=\gamma^{-1}(1-\beta \cos \vartheta)$ by adopting a jet
viewing angle $\vartheta = 30^\circ$ favoured in several recent works
\citep[{\em e.g.},][]{ng2004,komissarov2004,komissarov2010} and
assuming that the jet material moves at a speed similar to the range
of speeds observed in the wisps. This yields Doppler factors in a
1.4--1.8 range.  An upper limit on the $\mathcal{D}$ can also be
derived based upon the non-observation of an accompanying flare at
very high energy gamma-rays. The contemporaneous observations using
the VERITAS and MAGIC \citep{mariotti2010,ong2010} air Cherenkov
telescopes rule out any substantial deviations of the flux in the
energy regime between 100 GeV and TeV. Taking the relative flux errors
of roughly 10~\% as an upper limit on any additional flaring
component, we can estimate an upper limit on
$\mathcal{D}<\mathcal{O}(100)$. The actual upper limit depends
critically on the geometry. For the sake being conservative, we assume
a synchrotron self-Compton type scenario neglecting additional photon
fields.

These considerations (spectrum and size) {are consistent with the
hypothesis that the region C1 detected in the radio image is connected
with the high-energy flare in the Crab nebula. The unusually long
duration of higher flux state during this event may receive further
support from the detection of C2, which could be a second plasma
condensation ejected from the pulsar after the main flare in September
2010 and leading to emission enhancement in the high-energy regime.

The observed positional offset between the features C1 and A and the
lack of obvious optical/X-ray counterparts for the feature C2 raise
general questions of the relation between the emission detected in all
three bands and the nature of the flaring activity in the Crab
nebula. In the absence of better quality data, these questions remain
open. As a matter of speculation, one can suggest that the flaring
activity may be occurring in a relatively thin layer close to the jet
boundary, thus highlighting the jet edges where the path length
through the emitting material is larger (indeed, the locations of the
features A and C can also be reconciled with the same jet expansion as
derived above for C1 and C2).  Such an activity may originate, for
instance, from the magnetic reconnection or interaction of the jet
with the ambient medium.  However, in the absence of more detailed
observational data, it is clearly not feasible to provide a reliable
judgement on this matter.

In this respect, we also would like to emphasise that further
improvement of the quality of radio imaging on arcsecond scales (which
should be expected after the e-MERLIN becomes fully operational) would
provide much better reliability for imaging of such extended and
complicated objects as the central regions of the Crab
nebula. Therefore, engaging in a detailed monitoring program of the
Crab nebula with the EVN and e-MERLIN combined would be a highly
desirable option for a followup of any future strong flaring event in
this object.

\acknowledgements

APL acknowledges support from the Collaborative Research Center SFB
676 (Sonderforschungsbereich) ``Particle, Strings, and the Early
Universe'' funded by the German Research Society (Deutsche
Forschungsgemeinschaft). We are grateful to Michael Bietenholz for
providing a VLA image of the Crab Nebula. We thank the anonymous referee for 
constructive and helpful comments on the manuscript.

\bibliographystyle{aa} 

\bibliography{crab}

\begin{thebibliography}{48}
\expandafter\ifx\csname natexlab\endcsname\relax\def\natexlab#1{#1}\fi

\bibitem[{{Abdo} {et~al.}(2011){Abdo}, {Ackermann}, {Ajello}, {Allafort},
  {Baldini}, {Ballet}, {Barbiellini}, {Bastieri}, {Bechtol}, {Bellazzini},
  {Berenji}, {Blandford}, {Bloom}, {Bonamente}, {Borgland}, {Bouvier},
  {Brandt}, {Bregeon}, {Brez}, {Brigida}, {Bruel}, {Buehler}, {Buson},
  {Caliandro}, {Cameron}, {Cannon}, {Caraveo}, {Casandjian}, {{\c C}elik},
  {Charles}, {Chekhtman}, {Cheung}, {Chiang}, {Ciprini}, {Claus},
  {Cohen-Tanugi}, {Costamante}, {Cutini}, {D'Ammando}, {Dermer}, {de Angelis},
  {de Luca}, {de Palma}, {Digel}, {do Couto e Silva}, {Drell}, {Drlica-Wagner},
  {Dubois}, {Dumora}, {Favuzzi}, {Fegan}, {Ferrara}, {Focke}, {Fortin},
  {Frailis}, {Fukazawa}, {Funk}, {Fusco}, {Gargano}, {Gasparrini}, {Gehrels},
  {Germani}, {Giglietto}, {Giordano}, {Giroletti}, {Glanzman}, {Godfrey},
  {Grenier}, {Grondin}, {Grove}, {Guiriec}, {Hadasch}, {Hanabata}, {Harding},
  {Hayashi}, {Hayashida}, {Hays}, {Horan}, {Itoh}, {J{\'o}hannesson},
  {Johnson}, {Johnson}, {Khangulyan}, {Kamae}, {Katagiri}, {Kataoka}, {Kerr},
  {Kn{\"o}dlseder}, {Kuss}, {Lande}, {Latronico}, {Lee}, {Lemoine-Goumard},
  {Longo}, {Loparco}, {Lubrano}, {Madejski}, {Makeev}, {Marelli}, {Mazziotta},
  {McEnery}, {Michelson}, {Mitthumsiri}, {Mizuno}, {Moiseev}, {Monte},
  {Monzani}, {Morselli}, {Moskalenko}, {Murgia}, {Nakamori}, {Naumann-Godo},
  {Nolan}, {Norris}, {Nuss}, {Ohsugi}, {Okumura}, {Omodei}, {Ormes}, {Ozaki},
  {Paneque}, {Parent}, {Pelassa}, {Pepe}, {Pesce-Rollins}, {Pierbattista},
  {Piron}, {Porter}, {Rain{\`o}}, {Rando}, {Ray}, {Razzano}, {Reimer},
  {Reimer}, {Reposeur}, {Ritz}, {Romani}, {Sadrozinski}, {Sanchez},
  {Parkinson}, {Scargle}, {Schalk}, {Sgr{\`o}}, {Siskind}, {Smith}, {Spandre},
  {Spinelli}, {Strickman}, {Suson}, {Takahashi}, {Takahashi}, {Tanaka},
  {Thayer}, {Thompson}, {Tibaldo}, {Torres}, {Tosti}, {Tramacere}, {Troja},
  {Uchiyama}, {Vandenbroucke}, {Vasileiou}, {Vianello}, {Vitale}, {Wang},
  {Wood}, {Yang}, \& {Ziegler}}]{abdo2011}
{Abdo}, A.~A., {Ackermann}, M., {Ajello}, M., {et~al.} 2011, Science, 331, 739

\bibitem[{{Arendt} \& {Eilek}(2002)}]{arendt2002}
{Arendt}, Jr., P.~N. \& {Eilek}, J.~A. 2002, \apj, 581, 451

\bibitem[{{Arons} \& {Tavani}(1994)}]{arons1994}
{Arons}, J. \& {Tavani}, M. 1994, \apjs, 90, 797

\bibitem[{{Baars} {et~al.}(1977){Baars}, {Genzel}, {Pauliny-Toth}, \&
  {Witzel}}]{baars1977}
{Baars}, J.~W.~M., {Genzel}, R., {Pauliny-Toth}, I.~I.~K., \& {Witzel}, A.
  1977, \aap, 61, 99

\bibitem[{{Balbo} {et~al.}(2011){Balbo}, {Walter}, {Ferrigno}, \&
  {Bordas}}]{balbo2011}
{Balbo}, M., {Walter}, R., {Ferrigno}, C., \& {Bordas}, P. 2011, \aap, 527, L4+

\bibitem[{{Begelman}(1999)}]{begelman1999}
{Begelman}, M.~C. 1999, \apj, 512, 755

\bibitem[{{Bietenholz} {et~al.}(2004){Bietenholz}, {Hester}, {Frail}, \&
  {Bartel}}]{bietenholz2004}
{Bietenholz}, M.~F., {Hester}, J.~J., {Frail}, D.~A., \& {Bartel}, N. 2004,
  \apj, 615, 794

\bibitem[{{Bietenholz} {et~al.}(1997){Bietenholz}, {Kassim}, {Frail}, {Perley},
  {Erickson}, \& {Hajian}}]{bietenholz1997}
{Bietenholz}, M.~F., {Kassim}, N., {Frail}, D.~A., {et~al.} 1997, \apj, 490,
  291

\bibitem[{{Caraveo} {et~al.}(2010){Caraveo}, {de Luca}, {Mignani}, {Salvetti},
  {Bignami}, {Tavani}, {Costa}, {Pellizzoni}, {Buehler}, {D'Ammando}, {Hays},
  {Tennant}, {Weisskopf}, \& {Horns}}]{caraveo2010}
{Caraveo}, P., {de Luca}, A., {Mignani}, R., {et~al.} 2010, The Astronomer's
  Telegram, 2903, 1

\bibitem[{{Chedia} {et~al.}(1997){Chedia}, {Lominadze}, {Machabeli},
  {McHedlishvili}, \& {Shapakidze}}]{chedia1997}
{Chedia}, O., {Lominadze}, J., {Machabeli}, G., {McHedlishvili}, G., \&
  {Shapakidze}, D. 1997, \apj, 479, 313

\bibitem[{{Cheng} {et~al.}(2000){Cheng}, {Ruderman}, \& {Zhang}}]{cheng2000}
{Cheng}, K.~S., {Ruderman}, M., \& {Zhang}, L. 2000, \apj, 537, 964

\bibitem[{{Evangelista} {et~al.}(2010){Evangelista}, {Campana}, {Capalbi},
  {Mangano}, {Costa}, {Donnarumma}, {Feroci}, {Pacciani}, {Del Monte},
  {Lazzarotto}, {Soffitta}, {Lapshov}, {Tavani}, {Striani}, {Pittori},
  {Verrecchia}, {Romano}, {Bulgarelli}, {Gianotti}, {Trifoglio}, {Argan},
  {Trois}, {de Paris}, {Vittorini}, {Sabatini}, {Piano}, {D'Ammando}, {Chen},
  {Giuliani}, {Marisaldi}, {Di Cocco}, {Labanti}, {Fuschino}, {Galli},
  {Caraveo}, {Mereghetti}, {Perotti}, {Pucella}, {Rapisarda}, {Vercellone},
  {Pellizzoni}, {Pilia}, {Barbiellini}, {Longo}, {Picozza}, {Morselli},
  {Prest}, {Lipari}, {Zanello}, {Cattaneo}, {Rappoldi}, {Giommi},
  {Santolamazza}, {Lucarelli}, \& {Salotti}}]{evangelista2010}
{Evangelista}, Y., {Campana}, R., {Capalbi}, M., {et~al.} 2010, The
  Astronomer's Telegram, 2866, 1

\bibitem[{{Fendt}(1997)}]{fendt1997}
{Fendt}, C. 1997, \aap, 323, 999

\bibitem[{{Fendt} \& {Memola}(2001)}]{fendt2001}
{Fendt}, C. \& {Memola}, E. 2001, \aap, 365, 631

\bibitem[{{Ferrigno} {et~al.}(2010{\natexlab{a}}){Ferrigno}, {Tennant},
  {Horns}, {Weisskopf}, {Neronov}, {Tavani}, {Costa}, \&
  {Caraveo}}]{ferrigno2010b}
{Ferrigno}, C., {Tennant}, A., {Horns}, D., {et~al.} 2010{\natexlab{a}}, The
  Astronomer's Telegram, 2994, 1

\bibitem[{{Ferrigno} {et~al.}(2010{\natexlab{b}}){Ferrigno}, {Walter}, {Bozzo},
  \& {Bordas}}]{ferrigno2010a}
{Ferrigno}, C., {Walter}, R., {Bozzo}, E., \& {Bordas}, P. 2010{\natexlab{b}},
  The Astronomer's Telegram, 2856, 1

\bibitem[{{Gallant} \& {Arons}(1994)}]{gallant1994}
{Gallant}, Y.~A. \& {Arons}, J. 1994, \apj, 435, 230

\bibitem[{{Guirado} {et~al.}(1995){Guirado}, {Marcaide}, {Elosegui}, {Ratner},
  {Shapiro}, {Eckart}, {Quirrenbach}, {Schalinski}, \& {Witzel}}]{guirado1995}
{Guirado}, J.~C., {Marcaide}, J.~M., {Elosegui}, P., {et~al.} 1995, \aap, 293,
  613

\bibitem[{{Han} \& {Tian}(1999)}]{han1999}
{Han}, J.~L. \& {Tian}, W.~W. 1999, \aaps, 136, 571

\bibitem[{{Hankins} \& {Eilek}(2007)}]{hankins2007}
{Hankins}, T.~H. \& {Eilek}, J.~A. 2007, \apj, 670, 693

\bibitem[{{Hester} {et~al.}(2002){Hester}, {Mori}, {Burrows}, {Gallagher},
  {Graham}, {Halverson}, {Kader}, {Michel}, \& {Scowen}}]{hester2002}
{Hester}, J.~J., {Mori}, K., {Burrows}, D., {et~al.} 2002, \apjl, 577, L49

\bibitem[{{Hester} {et~al.}(1995){Hester}, {Scowen}, {Sankrit}, {Burrows},
  {Gallagher}, {Holtzman}, {Watson}, {Trauger}, {Ballester}, {Casertano},
  {Clarke}, {Crisp}, {Evans}, {Griffiths}, {Hoessel}, {Krist}, {Lynds},
  {Mould}, {O'Neil}, {Stapelfeldt}, \& {Westphal}}]{hester1995}
{Hester}, J.~J., {Scowen}, P.~A., {Sankrit}, R., {et~al.} 1995, \apj, 448, 240

\bibitem[{{Hester} {et~al.}(1998){Hester}, {Stapelfeldt}, \&
  {Scowen}}]{hester1998}
{Hester}, J.~J., {Stapelfeldt}, K.~R., \& {Scowen}, P.~A. 1998, \aj, 116, 372

\bibitem[{{Hirotani}(2006)}]{hirotani2006}
{Hirotani}, K. 2006, \apj, 652, 1475

\bibitem[{{Kaplan} {et~al.}(2008){Kaplan}, {Chatterjee}, {Gaensler}, \&
  {Anderson}}]{kaplan2008}
{Kaplan}, D.~L., {Chatterjee}, S., {Gaensler}, B.~M., \& {Anderson}, J. 2008,
  \apj, 677, 1201

\bibitem[{{Kennel} \& {Coroniti}(1984)}]{kennel1984}
{Kennel}, C.~F. \& {Coroniti}, F.~V. 1984, \apj, 283, 710

\bibitem[{{Komissarov} \& {Lyubarsky}(2004)}]{komissarov2004}
{Komissarov}, S.~S. \& {Lyubarsky}, Y.~E. 2004, \mnras, 349, 779

\bibitem[{{Komissarov} \& {Lyutikov}(2010)}]{komissarov2010}
{Komissarov}, S.~S. \& {Lyutikov}, M. 2010, ArXiv e-prints

\bibitem[{{Kovalenko} {et~al.}(1994){Kovalenko}, {Pynzar'}, \&
  {Udal'Tsov}}]{kovalenko1994}
{Kovalenko}, A.~V., {Pynzar'}, A.~V., \& {Udal'Tsov}, V.~A. 1994, Astronomy
  Reports, 38, 78

\bibitem[{{Lobanov}(2005)}]{lobanov2005}
{Lobanov}, A.~P. 2005, arXiv:astro-ph/0503225

\bibitem[{{Lorimer} {et~al.}(1995){Lorimer}, {Yates}, {Lyne}, \&
  {Gould}}]{lorimer1995}
{Lorimer}, D.~R., {Yates}, J.~A., {Lyne}, A.~G., \& {Gould}, D.~M. 1995,
  \mnras, 273, 411

\bibitem[{{Lyutikov}(2003)}]{lyutikov2003}
{Lyutikov}, M. 2003, \mnras, 339, 623

\bibitem[{{Mariotti}(2010)}]{mariotti2010}
{Mariotti}, M. 2010, The Astronomer's Telegram, 2967, 1

\bibitem[{{Mart{\'{\i}}-Vidal} {et~al.}(2008){Mart{\'{\i}}-Vidal}, {Marcaide},
  {Guirado}, {P{\'e}rez-Torres}, \& {Ros}}]{martividal2008}
{Mart{\'{\i}}-Vidal}, I., {Marcaide}, J.~M., {Guirado}, J.~C.,
  {P{\'e}rez-Torres}, M.~A., \& {Ros}, E. 2008, \aap, 478, 267

\bibitem[{{Matveyenko}(1975)}]{matveyenko1975}
{Matveyenko}, L.~I. 1975, Sov. Astr. Lett., 7, 13

\bibitem[{{Matveyenko} \& {Kostenko}(1979)}]{matveyenko1979}
{Matveyenko}, L.~I. \& {Kostenko}, V.~I. 1979, Aust. J. Phys., 32, 105

\bibitem[{{Melatos} {et~al.}(2005){Melatos}, {Scheltus}, {Whiting},
  {Eikenberry}, {Romani}, {Rigaut}, {Spitkovsky}, {Arons}, \&
  {Payne}}]{melatos2005}
{Melatos}, A., {Scheltus}, D., {Whiting}, M.~T., {et~al.} 2005, \apj, 633, 931

\bibitem[{{Meyer} {et~al.}(2010){Meyer}, {Horns}, \& {Zechlin}}]{meyer2010}
{Meyer}, M., {Horns}, D., \& {Zechlin}, H. 2010, \aap, 523, A2+

\bibitem[{{Ng} \& {Romani}(2004)}]{ng2004}
{Ng}, C. \& {Romani}, R.~W. 2004, \apj, 601, 479

\bibitem[{{Ng} \& {Romani}(2006)}]{ng2006}
{Ng}, C. \& {Romani}, R.~W. 2006, \apj, 644, 445

\bibitem[{{Ong}(2010)}]{ong2010}
{Ong}, R.~A. 2010, The Astronomer's Telegram, 2968, 1

\bibitem[{{Petrov} {et~al.}(2008){Petrov}, {Kovalev}, {Fomalont}, \&
  {Gordon}}]{petrov2008}
{Petrov}, L., {Kovalev}, Y.~Y., {Fomalont}, E.~B., \& {Gordon}, D. 2008, \aj,
  136, 580

\bibitem[{{Rees} \& {Gunn}(1974)}]{rees1974}
{Rees}, M.~J. \& {Gunn}, J.~E. 1974, \mnras, 167, 1

\bibitem[{{Spitkovsky} \& {Arons}(2004)}]{spitkovsky2004}
{Spitkovsky}, A. \& {Arons}, J. 2004, \apj, 603, 669

\bibitem[{{Tanvir} {et~al.}(1997){Tanvir}, {Thomson}, \&
  {Tsikarishvili}}]{tanvir1997}
{Tanvir}, N.~R., {Thomson}, R.~C., \& {Tsikarishvili}, E.~G. 1997, \na, 1, 311

\bibitem[{{Tavani} {et~al.}(2011){Tavani}, {Bulgarelli}, {Vittorini},
  {Pellizzoni}, {Striani}, {Caraveo}, {Weisskopf}, {Tennant}, {Pucella},
  {Trois}, {Costa}, {Evangelista}, {Pittori}, {Verrecchia}, {Del Monte},
  {Campana}, {Pilia}, {De Luca}, {Donnarumma}, {Horns}, {Ferrigno}, {Heinke},
  {Trifoglio}, {Gianotti}, {Vercellone}, {Argan}, {Barbiellini}, {Cattaneo},
  {Chen}, {Contessi}, {D'Ammando}, {DeParis}, {Di Cocco}, {Di Persio},
  {Feroci}, {Ferrari}, {Galli}, {Giuliani}, {Giusti}, {Labanti}, {Lapshov},
  {Lazzarotto}, {Lipari}, {Longo}, {Fuschino}, {Marisaldi}, {Mereghetti},
  {Morelli}, {Moretti}, {Morselli}, {Pacciani}, {Perotti}, {Piano}, {Picozza},
  {Prest}, {Rapisarda}, {Rappoldi}, {Rubini}, {Sabatini}, {Soffitta},
  {Vallazza}, {Zambra}, {Zanello}, {Lucarelli}, {Santolamazza}, {Giommi},
  {Salotti}, \& {Bignami}}]{tavani2011}
{Tavani}, M., {Bulgarelli}, A., {Vittorini}, V., {et~al.} 2011, Science, 331,
  736

\bibitem[{{Taylor} {et~al.}(1993){Taylor}, {Manchester}, \&
  {Lyne}}]{taylor1993}
{Taylor}, J.~H., {Manchester}, R.~N., \& {Lyne}, A.~G. 1993, \apjs, 88, 529

\bibitem[{{Tennant} {et~al.}(2010){Tennant}, {Caraveo}, {Costa}, {Evangelista},
  {Ferrigno}, {Heinke}, {Pellizzoni}, {Tavani}, \& {Weisskopf}}]{tennant2010}
{Tennant}, A., {Caraveo}, P., {Costa}, E., {et~al.} 2010, The Astronomer's
  Telegram, 2882, 1

\end{thebibliography}

\end{document}